\documentclass[12pt]{article}
\usepackage{amsmath,amssymb,amsfonts,epsfig}
\usepackage{graphics}
\usepackage{amssymb,amsmath}
\usepackage{cite}
\newcommand{\ct}{\cite}
\newcommand{\lb}{\label}

\newcommand{\bc}{\begin{center}}
\newcommand{\ec}{\end{center}}
\newcommand{\bd}{\begin{displaymath}}
\newcommand{\ed}{\end{displaymath}}
\newcommand{\be}{\begin{equation}}
\newcommand{\ee}{\end{equation}}
\newcommand{\ba}{\begin{array}}
\newcommand{\ea}{\end{array}}
\newcommand{\bea}{\begin{eqnarray}}
\newcommand{\eea}{\end{eqnarray}}
\newcommand{\bt}{\begin{tabular}}
\newcommand{\et}{\end{tabular}}

\newcommand{\bp}{\begin{picture}}
\newcommand{\ep}{\end{picture}}
\newcommand{\bfi}{\begin{figure}}
\newcommand{\efi}{\end{figure}}

\begin{document}

\begin{titlepage}

\begin{center}
{{\Large {\bf  {Cosmological Constant in a Model\\ with
Superstring-Inspired $E_6$ Unification\\\vskip0.2cm and Shadow
$\theta$-Particles}
}}}%
\vspace*{10mm}


{\bf \large{ C. R. Das $^{a}$ ,  L. V. Laperashvili $^{b}$  and A.
Tureanu $^{c}$}}

\end{center}
\begin{center}
\vspace*{0.4cm} {\it { $ ^a$\large Centre for Theoretical Particle
Physics, Technical University of Lisbon, \\ Avenida Rovisco Pais, 1
1049-001 Lisbon,
Portugal\\
$^b$ The Institute of Theoretical and Experimental Physics,\\
Bolshaya Cheremushkinskaya, 25, 117218 Moscow, Russia \\
$^c$ Department of Physics, University of Helsinki\\ and Helsinki
Institute of Physics, P.O.Box 64, FIN-00014 Helsinki, Finland}}\\
\vskip 0.2cm \centerline{\tt crdas@cftp.ist.utl.pt,laper@itep.ru,
anca.tureanu@helsinki.fi}

\vspace*{1.0cm}
\end{center}

\begin{center}{\bf Abstract}\end{center}
\begin{quote}
We have developed a concept of parallel existence of the ordinary
(O) and mirror (M), or shadow (Sh) worlds. In the first part of the
paper we consider a mirror world with broken mirror parity and the
breaking $E_6\to SU(3)^3$ in both worlds. We show that in this case
the evolutions of coupling constants in the O- and M-worlds are not
identical, having different parameters for similar evolutions. $E_6$
unification, inspired by superstring theory, restores the broken
mirror parity at the scale $\sim 10^{18}$ GeV. With the aim to
explain the tiny cosmological constant, in the second part we
consider the breakings: $E_6 \to SO(10)\times U(1)_Z$ -- in the
O-world, and $E'_6 \to SU(6)'\times SU(2)'_{\theta}$ -- in the
Sh-world. We assume the existence of shadow $\theta$-particles and
the low energy symmetry group $SU(3)'_C\times SU(2)'_L\times
SU(2)'_{\theta}\times U(1)'_Y$ in the shadow world, instead of the
Standard Model. The additional non-Abelian $SU(2)'_{\theta}$ group
with massless gauge fields, ''thetons", has a macroscopic confinement
radius $1/\Lambda'_{\theta}$. The assumption that
$\Lambda'_{\theta}\approx 2.3\cdot 10^{-3}$ eV explains the tiny
cosmological constant given by recent astrophysical measurements. In
this way the present work opens the possibility to specify a grand
unification group, such as $E_6$, from Cosmology.

\end{quote}
\vspace*{1.0cm}

\begin{flushright}\emph{Dedicated to the Memory of Kazuhiko Nishijima,\\
the founder of the concept of Shadow
Universe\cite{nishijima}.}\end{flushright}

\end{titlepage}



\section{Introduction}

Modern models for Dark Energy (DE) and Dark Matter (DM) are based on
precise measurements in cosmological and astrophysical observations
\ct{1,2,3,4,5}.

Supernovae observations at redshifts $1.25 \le z \le 1.7$ by the
Supernovae Legacy Survey (SNLS), cosmic microwave background
(CMB), cluster data and baryon acoustic oscillations by the Sloan
Digital Sky Survey (SDSS) fit the equation of state for DE: $w =
p/\rho$ with constant $w$, which is given by Ref. \cite{5}:
\be w = -1.023 \pm 0.090 \pm 0.054. \lb{1} \ee
The cosmological constant ($CC$) is given by the vacuum energy
density of the Universe \ct{1,2,3,4,5}:
\be CC = \rho_{vac}\approx (3\times 10^{-3}\,\,{\mbox{eV}})^4.
\lb{2} \ee

The result $w\approx - 1$, given by Eq. (\ref{1}), is consistent
with the present model of accelerating Universe \ct{6,7,8,9,10}
(see also reviews \ct{11,12,13}), dominated by a tiny cosmological
constant and Cold Dark Matter (CDM) -- this is the so-called
$\Lambda CDM$ scenario \ct{14}.

The present paper is devoted to the problem of cosmological
constant: why it is extremely small. This study develops some ideas
considered in Refs.\ct{15,16,17,18,19,20}, but leads to a new
interpretation of the possible structure of the Universe having such
a tiny $CC$.

Our model is based on the following assumptions:

$\bullet$ Grand Unified Theory (GUT) is inspired by the
superstring theory \ct{21,22,23,24,25,26}, which predicts $E_6$
unification in the 4-dimensional space \ct{26}, occurring at the
high energy scale $M_{E_6}\approx 10^{18}$ GeV.

$\bullet$ There exists a Mirror World (MW) \ct{27,28}, which is a
duplication of our Ordinary World (OW), or Shadow World (ShW)
\ct{nishijima,okun-pomer} (hidden sector \ct{29}) , which is not
identical with the O-world, having different symmetry groups.

$\bullet$ The M-world with broken mirror parity (MP) \ct{30,31,32},
or Shadow world \ct{15,16,17,18,19,20}, describes DE and DM.

$\bullet$ We assume that $E_6$ unification restores mirror parity at
high energies $\approx 10^{18}$ GeV (and at the early stage of the
Universe). Then the Mirror World exists and the group of symmetry of
the Universe is $E_6\times E'_6$ (the superscript 'prime' denotes
the M-world).

In this paper we consider two models with $E_6$ unification, one
with mirror world and one with shadow world. It is well known (see,
for example, Ref. \cite{46}) that there are three schemes of
breaking the $E_6$ group:
\begin{eqnarray}
i)\,\,\, E_6 &\to& SU(3)_1\times SU(3)_2\times SU(3)_3,\label{break1}\\
ii)\,\,\, E_6&\to& SO(10)\times U(1),\label{break2}\\
iii)\,\,\, E_6&\to& SU(6)\times SU(2)\label{break3}.
\end{eqnarray}
In the first case, we consider the possibility of the breaking
$$ E_6\to SU(3)_C\times SU(3)_L\times SU(3)_R  $$
in both O- and M-worlds, with broken mirror parity. The model has
the merit of an attractive simplicity. We should emphasize that this
breaking scheme is the only one which enables us to obtain the $E_6$
unification of the O- and M- worlds below the Planck scale and with
plausible values for the SUSY and seesaw scales. It is quite
impossible to obtain the same $E_6$ unification in the O- and
M-worlds if we have the same breakings $ii)$ or $iii)$ in both
worlds, with broken mirror parity. However, with this model we are
unable to explain the tiny $CC$ (\ref{2})\, given by astrophysical
measurements, because in this case we have in the low-energy limit
the SM in both worlds, which forbids a large confinement radius
(i.e. small scale) of any interaction. In the second case, we assume
different breakings of the $E_6$ unification in the O- and
Sh-worlds:
$$ E_6 \to SO(10)\times U(1), $$
$$ E'_6 \to SU(6)'\times SU(2)', $$
thus being able to explain the small value of the $CC$, due to the
additional $SU(2)'$ gauge symmetry group appearing at low energies
in the Sh-world, which has a large confinement radius. We should
point out that there are no other possibilities of breaking the
$E_6$ group, except the breakings \eqref{break1}-\eqref{break3}.

In the present paper we consider the idea of the existence of
theta-particles, developed by Okun \ct{33,34}\footnote{We are
grateful to M. Yu. Khlopov for this information.}. In those works it
was suggested the hypothesis that in Nature there exists the
symmetry group
\be SU(3)_C\times SU(2)_L\times U(1)_Y\times SU(2)_{\theta}\,,
\lb{6} \ee
i.e. with an additional non-Abelian $SU(2)_{\theta}$ group whose
gauge fields are neutral, massless vector particles -- thetons.
These thetons have a macroscopic confinement radius
$1/\Lambda_{\theta}$. Here we assume that such a group of symmetry
exists in the Shadow World at low energy with $\Lambda'_{\theta}\sim
10^{-3}$ eV and provides the tiny cosmological constant.

\section{Superstring theory and $E_6$ unification}

\subsection{Superstring theory}

Superstring theory \ct{21,22,23,24,25,26} is a paramount candidate
for the unification of all fundamental gauge interactions with
gravity. Superstrings are free of gravitational and Yang-Mills
anomalies if the gauge group of symmetry is $SO(32)$ or $E_8\times
E_8$.

The 'heterotic' superstring theory $E_8\times E'_8$ was suggested as
a more realistic model for unification \ct{23,24}. This
ten-dimensional Yang-Mills theory can undergo spontaneous
compactification. The integration over six compactified dimensions
of the $E_8$ superstring theory leads to the effective theory with
the $E_6$ unification in the four-dimensional space \ct{26}. Among
hundreds of papers devoted to the $E_6$ unification, we would like
to single out Refs. \cite{7a,7b,7c,7d,7e,7g,7f,7ff}.


\subsection{The group $E_6$}

Three 27-plets of $E_6$ contain three families of quarks and
leptons, including right-handed neutrinos $N_i^c$ ($i=1,2,3$ is the
index of generations). Matter fields (quarks and leptons) of the
fundamental 27-representation of the flipped $E_6$ decompose under
$SU(5)\times U(1)_X$ subgroup as follows:
\be
        27 \to (10,1)  + (\bar 5, -3)+
          (5,-2)+ (\bar 5,2)  + (1,5) + (1,0).          \lb{1a}
\ee
The first and second numbers in the brackets of Eq. (\ref{1a})
correspond to the dimensions of the $SU(5)$ representation and to
the $U(1)_X$ charges, respectively. The Standard Model (SM) family
which contains the doublets of left-handed quarks $Q$ and leptons
$L$, right-handed up and down quarks $u^c$, $d^c$, also $e^c$ and
right-handed neutrino $N^c$ belongs to the $(10,1) + (\bar 5,-3) +
(1,5)$ representations of the flipped $SU(5)\times U(1)_X$. These
representations decompose under the groups with the breakings
\be SU(5)\times U(1)_X \to SU(3)_C\times SU(2)_L\times U(1)_Z\times
U(1)_X.\lb{3a} \ee

Then, for the decomposition (\ref{3a}), we have the following
assignments of particles:
\bea
       (10,1) \to Q = &\left(\begin{array}{c}u\\
                                          d \end{array}\right) &\sim
                         \left(3,2,\frac 16,1\right),\nonumber\\
&d^{\rm\bf c} &\sim \left(\bar3,1,-\frac 23,1\right),\nonumber\\
&N^{\rm\bf c} &\sim \left(1,1,1,1\right).       \lb{4a}\\
(\bar 5,-3) \to &u^{\rm\bf c}&\sim \left(\bar 3,1,\frac
13,-3\right),
\nonumber\\
L = &\left(\begin{array}{c}e\\
                                             \nu \end{array}\right) &\sim
                         \left(1,2,-\frac 12,-3\right),               \lb{5a}\\
(1,5) \to &e^{\rm\bf c} &\sim \left(1,1,1,5\right).\lb{6a} \eea
The remaining representations in (\ref{3a}) decompose as follows:
\bea
        (5,-2) \to& D&\sim \left(3,1,-\frac 13,-2\right),\nonumber\\
                   h = &\left(\begin{array}{c}h^+\\
                                               h^0 \end{array}\right) &\sim
                         \left(1,2,\frac 12,-2\right).
                                                              \lb{7a}\\
    (\bar 5,2) \to &D^{\rm\bf c} &\sim \left(\bar 3,1,\frac 13,2\right),
\nonumber\\
                     h^{\rm\bf c} = &\left(\begin{array}{c}h^0\\
                                               h^- \end{array}\right) &\sim
                         \left(1,2,-\frac 12,2\right).              \lb{8a}
\eea
The light Higgs doublets are accompanied by the heavy coloured Higgs
triplets $D,D^{\rm\bf c}$ which are absent in the SM. The singlet
field $S$ is represented by (1,0): \be
       (1,0) \to S.               \lb{9a}
\ee
Let us remark that the flipping of our $SU(5)$,
\be
       d^{\rm\bf c} \leftrightarrow u^{\rm\bf c},\quad
N^{\rm\bf c}\leftrightarrow e^{\rm\bf c}, \lb{10a} \ee
differentiates this group of symmetry from the standard
Georgi-Glashow $SU(5)$ \ct{27a}.

\section{$E_6$ unification in ordinary and mirror world}

In this Section we consider the hypothesis that there exists in
Nature a mirror world, parallel to our ordinary  world \ct{27,28}
(see also Refs.
\cite{35,5a,5b,5c,5d,5e,5g,5i,5j,5k,36,601,602,603,37}). This
M-world is a mirror copy of the O-world and contains the same
particles and types of interactions as our visible world. The
observable elementary particles of our O-world have the left-handed
(V-A) weak interactions which violate P-parity. If a hidden mirror
M-world exists, then mirror particles participate in the
right-handed (V+A) weak interactions and have the opposite
chirality.

Lee and Yang were the first \ct{27} to suggest such a duplication
of the worlds, which restores the left-right symmetry of Nature.
They introduced a concept of right-handed particles, but their
R-world was not hidden. The term 'Mirror Matter' was introduced by
Kobzarev, Okun and Pomeranchuk \ct{28}. They suggested the 'Mirror
World' as the hidden sector of our Universe, which interacts with
the ordinary (visible) world only via gravity or another very weak
interaction. They have investigated a variety of phenomenological
implications of such parallel worlds (for recent comprehensive
reviews on mirror particles and mirror matter, see Refs.
\cite{okun-rev,blin-rev}).

We have assumed that at very high energies $\sim 10^{18}$ GeV, there
exists $E_6$ unification, predicted by superstring theory, in both
O- and M-world. In this case, mirror parity (MP) is restored and we
have the group of symmetry $E_6\times E'_6$.

\subsection{Particle content in the ordinary and mirror worlds}

At low energies we can describe the ordinary and mirror worlds by a
minimal symmetry $$G_{SM}\times G'_{SM}, \quad
$$ where
$$G_{SM} = SU(3)_C\times SU(2)_L\times U(1)_Y$$ stands
for the Standard Model (SM) of observable particles: three
generations of quarks and leptons and the Higgs boson. Then
$$G'_{SM} = SU(3)'_C\times SU(2)'_L\times U(1)'_Y$$ is its mirror
gauge counterpart having three generations of mirror quarks and
leptons and the mirror Higgs boson. The M-particles are singlets of
$G_{SM}$ and the O-particles are singlets of $G'_{SM}$. {\it These
different O- and M-worlds are coupled only by gravity, or possibly
by another very weak interaction.} Including the Higgs bosons
$\Phi$, we have the following SM content of the O-world:
$$\rm{
L-set}: {\quad (u,d,e,\nu,\tilde u,\tilde d,\tilde e,\tilde
N)_L\,,\Phi_u,\,\Phi_d} ;$$
$$ \rm { \tilde R-set}: { \quad (\tilde
u,\tilde d,\tilde e,\tilde \nu,u,d,e,N)_R\,,\tilde \Phi_u,\,\tilde
\Phi_d;}$$
with the antiparticle fields: ${\tilde \Phi_ {u,d} =
\Phi^*_{u,d},}\,\,$ ${ \tilde \psi_R = C\gamma_0\psi_L^*\,\,}$ and
${\tilde \psi_L = C\gamma_0\psi_R^*.}$

Considering the minimal symmetry $G_{SM}\times G'_{SM}$, we have the
following particle content in the M-sector:
$$ \rm{ L'-set}: \quad \Large
(u',d',e',\nu',\tilde u',\tilde d',\tilde e',\tilde
N')_L\,,\Phi'_u,\,\Phi'_d ;$$
$${\rm
 \tilde R'-set}: {\quad (\tilde u',\tilde d',\tilde e',\tilde
\nu',u',d',e',N')_R\,,\tilde \Phi'_u,\,\tilde \Phi'_d.}$$
In general, we can consider a supersymmetric theory when $ G\times
G'$ contains the grand unification groups $SU(5)\times SU(5)'$,
$SO(10)\times SO(10)',\,\,$ $E_6\times E_6'$ etc.

\subsection{Mirror world with broken mirror parity}

In the general case, mirror parity (MP) is not conserved, and the
ordinary and mirror worlds are not identical
\ct{30,31,32,36,601,602,603,37} in the sense that, although the
chain of breakings of the gauge groups is the same in both worlds,
the energy scales at which these breakings take place are different.

If the O- and M-sectors are described by the minimal symmetry group
$$G_{SM}\times G'_{SM} $$
with the Higgs doublets $\Phi$ and $\Phi'$, respectively, then in
the case of non-conserved MP the VEVs of $\Phi$ and $\Phi'$ are not
equal: $v\neq v'$. In accord with Refs.
\cite{30,31,32,36,601,602,603,37}, we assume that
$$v'>>v$$ and introduce the parameter characterizing the
violation of MP: \be
       \zeta = \frac{v'}{v} >> 1. \lb{7}  \ee
Then the masses of fermions and massive bosons in the mirror world
are scaled up by the factor $\zeta$ with respect to the masses of
their counterparts in the ordinary world:
\begin{eqnarray}
               m'_{q',l'} &=& \zeta m_{q,l}, \lb{8} \\
                 M'_{W',Z',\Phi'} &=& \zeta M_{W,Z,\Phi}, \lb{9}
                 \end{eqnarray}
while photons and gluons remain massless in both worlds.

Let us consider now the expressions for the running of the inverse
coupling constants,
\begin{eqnarray}
    \alpha_i^{-1}(\mu) &=& \frac{b_i}{2\pi}\ln
    \frac{\mu}{\Lambda_i},\ \quad \mbox{in the O-world;}  \lb{10}\\
{\alpha'}_i^{-1}(\mu) &=& \frac{b'_i}{2\pi}\ln
    \frac{\mu}{\Lambda'_i}, \ \quad \mbox{in the M-world}.       \lb{11}
    \end{eqnarray}
Here $i=1,2,3$ correspond to $U(1),\, SU(2)$ and $SU(3)$ groups of
the SM (or SM$'$). A big difference between the electroweak scales
$v$ and $v'$ will not cause the same difference between the scales
$\Lambda_i$ and $\Lambda'_i.$ Hence, \be \Lambda'_i = \xi \Lambda_i.
\lb{12} \ee The value of $\,\,\zeta \,\,$ was estimated by
astrophysical implications \ct{30,31,32} which led to
\be \zeta\approx 30. \lb{11s}  \ee
In general, $\zeta$ can be considered in the range 10-100.\\

\subsection{Seesaw scale in the ordinary and mirror
worlds}

In the language of neutrino physics, the O-neutrinos
$\nu_e,\,\,\nu_{\mu},\,\,\nu_{\tau}$ are active neutrinos, while the
M-neutrinos $\nu'_e,\,\,\nu'_{\mu},\,\,\nu'_{\tau}$ are sterile
neutrinos. The model \ct{30,31,32,36,601,602,603,37} provides a
simple explanation of why sterile neutrinos could be light, and
could have significant mixing with the active neutrinos.

If MP is conserved ($\zeta = 1$), then the neutrinos of the two
sectors are strongly mixed. But it seems that the situation with the
present experimental and cosmological limits on the active-sterile
neutrino mixing do not confirm this result. If instead MP is
spontaneously broken, and $\zeta >> 1$, then the active-sterile
mixing angles should be small:
\be \theta_{\nu\nu'}\sim \frac 1{\zeta}. \lb{12s} \ee As a result,
we have the following relation between the masses of the light
left-handed neutrinos:
\be
  m'_{\nu}\approx \zeta^2 m_{\nu}. \lb{13s} \ee

In the context of the SM, in addition to the fermions with non-zero
gauge charges, one introduces also the gauge singlets, the so-called
right-handed neutrinos $N_a$ with large Majorana mass terms.
According to Refs.~\ct{30,31,32,36,601,602,603,37}, they have equal
masses in the O- and M-worlds:
\be M'_{\nu,a} = M_{\nu,a}. \lb{14s} \ee

Let us consider now the usual seesaw mechanism. Heavy right-handed
neutrinos are created at the seesaw scales $M_R$ in the O-world and
$M'_R$ in the M- or Sh-world. From the Lagrangian, considering the
Yukawa couplings identical in the two sectors, it follows that
\be m_{\nu}^{(')}=\frac {{v^{(')}}^2}{M_R^{(')}}, \lb{15s} \ee and
we immediately obtain the relations (\ref{13s}), with
\be M'_R = M_R \lb{16s}. \ee
Then we see that even in the model with broken mirror parity, we
have the same seesaw scales in the O- and M- or Sh-worlds.

\section{Model with breaking
$E_6 \to SU(3)_C\times SU(3)_L\times SU(3)_R$ in ordinary and mirror
worlds}

We assume that in the ordinary and mirror worlds there exists the
symmetry
$$G\times G'.$$ In the ordinary world, from the SM up to the
$E_6$ unification, $G$ represents the following chain of restoration
of symmetry groups:
$$  SU(3)_C\times SU(2)_L\times U(1)_Y $$ $$\to  [SU(3)_C\times
SU(2)_L\times U(1)_Y]_{{SUSY}}
$$ $$\to
 SU(3)_C\times SU(2)_L \times SU(2)_R\times U(1)_X\times
 U(1)_Z$$
 $$  \to  SU(3)_C\times SU(3)_L \times SU(2)_R\times
 U(1)_Z $$ \be  \to SU(3)_C\times SU(3)_L \times SU(3)_R \to
 E_6,  \lb{16} \ee
and in the mirror world $G'$ represents the symmetry group chain of
the same structure:
$$SU(3)'_C\times
SU(2)'_L\times U(1)'_Y $$ $$\to  [SU(3)'_C\times SU(2)'_L\times
U(1)'_Y]_{{SUSY}}
$$ $$\to
 SU(3)'_C\times SU(2)'_L \times SU(2)'_R\times U(1)'_X\times
 U(1)'_Z$$
 $$ \to  SU(3)'_C\times SU(3)'_L \times SU(2)'_R\times
 U(1)'_Z $$ \be \to SU(3)'_C\times SU(3)'_L \times SU(3)'_R \to
 E'_6. \lb{17} \ee
Also we have assumed that the $E_6$ unification, being the same in
the O- and M-worlds ($E_6 = E'_6$, which means the same unification
scales, $M_{E_6}=M'_{E_6}$, with the same
super-super-GUT\footnote{The prefix ''super" refers to higher
unification scales, in the order GUT, super-GUT, super-super-GUT
etc.} coupling constants, $g_{E_6}=g'_{E_6})$, restores the broken
mirror parity MP at the super-super-GUT-scale $M_{E_6}\sim
{10}^{18}$ GeV. At low energies we have the Standard Model in both
worlds,
         $G_{SM}\times G'_{SM},$
for which case the mirror world with broken mirror parity has been
studied in Refs. \cite{30,31,32,36,601,602,603,37}.

\subsection{Gauge coupling constant evolutions in the
O- and M-worlds} In this work we consider the running of all the
gauge coupling constants in the SM and its extensions which is well
described by the one-loop approximation of the renormalization group
equations (RGEs), since from the Electroweak (EW) scale up to the
Planck scale ($M_{Pl}$) all the non-Abelian gauge theories with rank
$r\ge 2$ appearing in our model are chosen to be asymptotically
free. With this aim we consider only the Higgs bosons belonging to
the $N+\bar N$ representations for $SO(N)$ or $SU(N)$ symmetry
breaking \cite{7f}.

For the energy scale $\mu \ge M_{{ren}}$, where $M_{{ren}}$ is the
renormalization scale, we have the following evolution for the
inverse coupling constants $\alpha_i^{-1}$ given by RGE in the
one-loop approximation:
 \be
 \alpha_i^{-1}(\mu) = \alpha_i^{-1}(M_{ren}) + \frac{b_i}{2\pi}t, \lb{18} \ee
where
 $$ \alpha_i =\frac {g^2_i}{4\pi} $$
and $g_i$ is the gauge coupling constant of the gauge group $G_i$.
Here
$$ t=\ln\left(\frac{\mu}{M_{ren}}\right).$$
The coefficients (slopes) $b_i$, describing the running of the
coupling constants with our choice of gauge groups and particle
content, are given in Table 1 (according to Refs.
\cite{19,7f,38,39}), for both O- and M-worlds.

\begin{table}
\begin{center}
{\scriptsize
\begin{tabular}{|r|c|c|c|c|c|}\hline
NonSUSY groups:&$SU(3)_C$              &$SU(2)_L$    &$U(1)_Y$                &                                     \\
$b_i$:         &$7$                    &$19/6$       &$-41/10$ &
\\\hline

SUSY groups:   &$SU(3)_C$              &$SU(2)_{L,R}$&$SU(2)_L\times SU(2)_R $&$U(1)_Y$                             \\
$b_i^{SUSY}$:  &$3$                    &$-1$         &$b_{22}=-2$             &$-33/5$                              \\

               &                       &             &                        &                                     \\
               &$SU(3)_C\times SU(3)_L$&$U(1)_X$     &$U(1)_Z$                &$SU(3)_C\times SU(3)_L\times SU(3)_R$\\
               &$b_{33}=9$             &$-33/5$      &$-33/5$                 &$b_{333}=21$                         \\\hline
\end{tabular}
} \caption{The coefficients $b_i$ in the O- and M-worlds.}
\end{center}
\end{table}

In the following, we shall use the fact that the running of the
coupling constants is the same in the O- and M-world, since the
gauge groups of symmetry and the particle content are the same in
the two worlds, but the scales of the gauge symmetry breakings are
different, due to the violation of mirror parity.

In our model we shall consider
\begin{equation}
\zeta=10.
\end{equation}

\subsection{Standard Model}\label{SM}

We start with the SM in the ordinary world:
$$ G_{{SM}} = SU(3)_C\times SU(2)_L\times
U(1)_Y$$
and SM$'$ in the mirror world: $$G'_{{SM'}} = SU(3)'_C\times
SU(2)'_L\times U(1)'_Y. $$
For compactness of notation, in the following we shall denote by
$\alpha^{(\prime)-1}$ the inverse of various coupling constants and
by $M^{(\prime)}_{ren}$ the various renormalization scales belonging
to either ordinary world (the non-primed symbols) or mirror world
(the primed symbols). For energy scales $\mu \ge M_t$ ($M_t$ -- top
quark mass) in the SM and $\mu \ge M'_t$ ($M'_t=\zeta M_t$ -- mirror
top quark mass) in the SM$'$ we have the following evolutions (RGEs)
\ct{38,39,40} for the inverse coupling constants
$\alpha_i^{(\prime)-1}$ ($i=1,2,3$ correspond to the
$U(1)^{(\prime)}$, $SU(2)^{(\prime)}_L$ and $SU(3)^{(\prime)}_C$
groups of the SM$^{(\prime)}$, respectively):
\be \alpha_1^{(\prime)-1}(t) =
\alpha_1^{(\prime)-1}(M^{(\prime)}_t) -
\frac{41}{20\pi}t^{(\prime)}, \lb{13} \ee
\be       \alpha_2^{(\prime)-1}(t) =
\alpha_2^{(\prime)-1}(M^{(\prime)}_t) +
\frac{19}{12\pi}t^{(\prime)},  \lb{14} \ee
 \be
\alpha_3^{(\prime)-1}(t) = \alpha_3^{(\prime)-1}(M^{(\prime)}_t) +
\frac{7}{2\pi}t^{(\prime)},  \lb{19} \ee
where
\be  \alpha_1^{-1}(M_t) = \alpha_1^{\prime-1}(M^{\prime}_t) =58.65
\pm 0.02, \lb{20} \ee
\be    \alpha_2^{-1}(M_t) =\alpha_2^{\prime-1}(M^{\prime}_t) = 29.95
\pm 0.02, \lb{21} \ee
\be \alpha_3^{-1}(M_t) = \alpha_3^{\prime-1}(M^{\prime}_t) =9.17 \pm
0.20, \lb{22} \ee
and the evolution variables are
$$
t=\ln\left(\frac{\mu}{M_t}\right)\ \ \mbox{and}\ \
t'=\ln\left(\frac{\mu}{M'_t}\right).$$
We have used the central value of the top quark mass (Particle Data
Group \ct{41}):
$$ M_t= 172.6\,\, {\mbox{GeV}},$$
implying, for $\zeta=10,$ $M'_t= 1.726\,\, {\mbox{TeV}}.$

In Eq.~(\ref{22}) the value of $\alpha_3^{-1}(M_t) = 9.17$
essentially depends on the value of $$ \alpha_3(M_Z)\equiv
\alpha_s(M_Z) = 0.118\pm 0.002$$ (see Particle Data Group
\ct{41}), where $M_Z$ is the mass of the $Z$-boson. The value of
$\alpha_3^{-1}(M_t)$ is given by the running of $
\alpha_3^{-1}(\mu)$ from $M_Z$ up to $M_t$, via the Higgs boson
mass $M_H$. Here we have used $M_H = 130\pm 15 \,\,\,\rm{GeV},$ in
accord with the observed experimental data at LEP2 and Tevatron.

If we assume now that for $\mu \le M_t$ in the O-world:
\be
    \alpha_2^{-1}(\mu) = \frac{b_2}{2\pi}\ln
    \frac{\mu}{\Lambda_2},  \lb{4a} \ee
and for $\mu \le M'_t$ in the M-world:
\be {\alpha'}_2^{-1}(\mu) = \frac{b_2}{2\pi}\ln
    \frac{\mu}{\Lambda'_2},        \lb{5a} \ee
with
\be \Lambda'_2 < \Lambda_2 \,\,\, (\xi < 1\ \mbox{in}\ \eqref{12}),
 \lb{6a} \ee
then we know that ${\alpha_2^{(\prime)-1}}(\mu)$ runs down to low
energies, but stops at some scale with coupling constant $
g^{(\prime)}_2$ corresponding to the Fermi constant $G_F$ of the
4-fermion weak interaction. This is a consequence of the existence
of the Higgs particles. As it was shown in Subsection 3.2, the SM
VEVs of the Higgs fields are: $\langle\Phi\rangle=v=246$ GeV and
$\langle\Phi'\rangle=v'=\zeta v$. Then the intermediate bosons
$W(W')$ and $Z(Z')$ acquire the masses $M_{W,Z}$ known in the SM and
$M'_{W',Z'}$ given by Eq.~(\ref{9}). As a result, an extremely large
confinement radius is absent. But if the intermediate bosons would
not acquire mass due to the Higgs mechanism, then their gauge
interaction would be characterized by a macroscopic radius of
confinement. In this case we could have a scale $\Lambda_2$ or
$\Lambda'_2$  at extremely low energies. However, we do not have
such an extremely small scale in the Standard Model.

\subsection{MSSM}

The Minimal Supersymmetric Standard Model (MSSM$^{(\prime)}$) (which
extends the conventional SM$^{(\prime)}$) gives the evolutions for
$\alpha_i^{(\prime)-1}(\mu)$ ($i=1,2,3\,\,$ correspond to the
$U(1)^{(\prime)}$, $SU(2)^{(\prime)}$, $SU(3)^{(\prime)}$ groups,
respectively) from the supersymmetric scale $M^{(\prime)}_{{SUSY}}$
up to the seesaw scale $M^{(\prime)}_R$, where the heavy (mirror)
right-handed neutrinos are produced. In MSSM$'$ the superpartners of
particles, i.e., ''sparticles", have in the M-world the masses
$\tilde m' = \zeta \tilde m.$ This means that the supersymmetry
breaking scale in the M-world is larger:
$$M'_{{SUSY}} = \zeta M_{{SUSY}}.$$
At the seesaw scale $M'_R = M_R,$ the mirror right-handed neutrinos
appear.

From \eqref{13}-\eqref{19}, one easily obtains:
\be \alpha_1^{-1}(M_{SUSY}) =\alpha_1^{'-1}(M'_{SUSY})=56.01 ,
\lb{23} \ee
\be \alpha_2^{-1}(M_{SUSY}) =\alpha_2^{'-1}(M'_{SUSY})=31.99
,\lb{24} \ee
\be \alpha_3^{-1}(M_{SUSY)} =\alpha_3^{'-1}(M'_{SUSY})=13.68\,,
\lb{25} \ee
for $M_{SUSY}=10\ \rm{TeV}$ and $M'_{SUSY}=\zeta M_{SUSY}=100\
\rm{TeV}$, when $\zeta=10$. Above these scales we have, according to
Table 1:
\be
      \alpha_1^{(\prime)-1}(t^{(\prime)}_s) = \alpha_1^{(\prime)-1}(M^{(\prime)}_{SUSY}) -
      \frac{33}{10\pi}t^{(\prime)}_s,  \lb{26} \ee
\be \alpha_2^{(\prime)-1}(t^{(\prime)}_s) =
\alpha_2^{(\prime)-1}(M^{(\prime)}_{SUSY}) -
\frac{1}{2\pi}t^{(\prime)}_s, \lb{27} \ee
\be \alpha_3^{(\prime)-1}(t^{(\prime)}_s) =
\alpha_3^{(\prime)-1}(M^{(\prime)}_{SUSY}) +
\frac{3}{2\pi}t^{(\prime)}_s, \lb{28} \ee
with the respective evolution parameters in the O- and M-worlds:
$$t_s=\ln\left(\frac{\mu}{M_{SUSY}}\right)\ \ \mbox{and}\ \
t'_s=\ln\left(\frac{\mu}{M'_{SUSY}}\right).$$

\subsection{Left-right symmetry}

We assume that the following supersymmetric left-right symmetry
originates in the O-world at the seesaw scale $M_R$ \ct{42,43,44}:
$$
 SU(3)_C\times SU(2)_L\times
SU(2)_R\times U(1)_X\times U(1)_Z, $$
and, correspondingly, in the M-world at the mirror seesaw scale
$M'_R=M_R$:
 $$ SU(3)'_C\times SU(2)'_L\times SU(2)'_R\times
U(1)'_X\times U(1)'_Z. $$

The following evolutions appear at the seesaw scale $M^{(\prime)}_R$
and take place for $\mu\ge M^{(\prime)}_R$:
\be \alpha_X^{(\prime)-1}(t^{(\prime)}_r) =
\alpha_X^{(\prime)-1}(M^{(\prime)}_{R}) -
      \frac{33}{10\pi}t^{(\prime)}_r,  \lb{29} \ee
\be
      \alpha_Z^{(\prime)-1}(t^{(\prime)}_r) = \alpha_Z^{(\prime)-1}(M^{(\prime)}_R) -
      \frac{33}{10\pi}t^{(\prime)}_r,  \lb{30} \ee
\be \alpha_{22}^{(\prime)-1}(t^{(\prime)}_r) =
\alpha_{22}^{(\prime)-1}(M^{(\prime)}_R) -
\frac{1}{\pi}t^{(\prime)}_r,\lb{31} \ee
with
\be t_r=\ln\left(\frac{\mu}{M_R}\right)\ \ \mbox{and}\ \
t'_r=\ln\left(\frac{\mu}{M'_R}\right).\lb{31a}\ee

\subsection{$SU(3)_C\times SU(3)_L$,
$SU(3)_C\times SU(3)_L\times SU(3)_R$ and $E_6$ unification}

The intersection of the evolutions (\ref{28}) and (\ref{31}) for the
O-world couplings gives the scale $M_{GUT}= 3.91\cdot 10^{15}$ GeV.
The evolution of the Abelian group $U(1)_X$: \be
      \alpha_X^{-1}(t_r) = \alpha_X^{-1}(M_{R}) -
      \frac{33}{10\pi}t_r,  \lb{32} \ee
which has appeared at the seesaw scale $M_R$, meets the point
$M_{GUT}$.

Then we have the evolution for $SU(3)_C\times SU(3)_L$ from
$M_{GUT}= 3.91\cdot 10^{15}$ GeV up to the super-GUT scale
$M_{SGUT}= 3\cdot 10^{18}$ GeV : \be \alpha_{33}^{-1}(t_g) =
\alpha_{33}^{-1}(M_{GUT}) + \frac{9}{2\pi}t_g, \lb{33} \ee
with the evolution parameter
$$t_g=\ln\left(\frac{\mu}{M_{GUT}}\right).$$
Here $M_{SGUT}$ is the scale of the $SU(3)_C\times SU(3)_L\times
SU(3)_R$-unification.

In our model, from $M_{SGUT}=3\cdot 10^{18}$ GeV up to
$M_{E_6}=5\cdot 10^{18}$ GeV we have the evolution for
$SU(3)_C\times SU(3)_L\times SU(3)_R$:
\be \alpha_{333}^{-1}(t_{sg}) = \alpha_{333}^{-1}(M_{SGUT}) +
\frac{21}{2\pi}t_{sg}, \lb{33a} \ee
with
$$t_{sg}=\ln\left(\frac{\mu}{M_{SGUT}}\right).$$
From $M_{SGUT}=3\cdot 10^{18}$ GeV  down to $M_R= 10^{12}$ GeV we
have the evolution of the Abelian  $U(1)_Z$-group:
\be
      \alpha_Z^{-1}(t_r) = \alpha_Z^{-1}(M_R) -
      \frac{33}{10\pi}t_r,  \lb{34} \ee
and down to $M_{GUT}$ -- the evolution of the non-asymptotically
free supersymmetric $SU(2)_R$-group:
\be  \alpha_{2R}^{-1}(t_r) = \alpha_{2R}^{-1}(M_R) -
\frac{1}{2\pi}t_r,  \lb{35} \ee with $t_r$ given by Eq.~(\ref{31a}).

\
\subsection{Mirror symmetries $SU(3)'_C\times SU(3)'_L$ and
$SU(3)'_C\times SU(3)'_L\times SU(3)'_R$}

The intersection of the evolutions (\ref{28}) and (\ref{31}) for
 the M-world couplings leads to the scale
$M'_{GUT}= 2.46\cdot 10^{16}$ GeV. The evolution (\ref{29}) of
$\alpha_X^{\prime-1}(t^{\prime}_r)$ which begins its running from
the seesaw scale $M'_R$, has its end at the point $M'_{GUT}$. Then
we have the evolution for $SU(3)'_C\times SU(3)'_L$ from $M'_{GUT}=
2.46\cdot 10^{16}$ GeV up to $M'_{SGUT}=10^{17}$ GeV (which is
arbitrarily chosen, because it is not given by the theory):
\be {\alpha'}_{33}^{-1}(t'_g) = {\alpha'}_{33}^{-1}(M'_{GUT}) +
\frac{9}{2\pi}t'_g, \lb{26a} \ee
with
$$t'_g=\ln\left(\frac{\mu}{M'_{GUT}}\right).$$
From $M'_{SGUT}= 10^{17}$ GeV  up to $M_{E_6}=M'_{E_6}=5\cdot
10^{18}$ GeV we have the evolution: \be
{\alpha'}_{333}^{-1}(t'_{sg}) = {\alpha'}_{333}^{-1}(M'_{SGUT}) +
\frac{21}{2\pi}t'_{sg} \lb{27a} \ee with the evolution parameter
$$t'_{sg}=\ln\left(\frac{\mu}{M'_{SGUT}}\right).$$ From
$M'_{SGUT}=10^{17}$ GeV  down to $M'_R=10^{12}$ GeV  we have the
evolutions: \be
      {\alpha'}_Z^{-1}(t'_r) = {\alpha'}_Z^{-1}(M'_R) -
      \frac{33}{10\pi}t'_r,  \lb{28a} \ee
\be      {\alpha'}_{2R}^{-1}(t'_r) = {\alpha'}_{2R}^{-1}(M'_R) -
\frac{1}{2\pi}t'_r.  \lb{29a} \ee
The total picture of the evolutions in the O- and M-worlds is
presented simultaneously in Fig. 1 for the case:
$$M_{SUSY}= 10\,\,\, {\mbox
{TeV}},$$ $$M_R=10^{12}\,\,\, {\mbox {GeV}},$$ which give:
$$M_{GUT}= 3.91\cdot 10^{15}\,\,\, {\mbox {GeV}}. $$
The value of the super-GUT scale in the O-world,
$$M_{SGUT}= 3\cdot 10^{18}\,\,\, {\mbox {GeV}} $$
depends of the choice of $M'_{SGUT}$.

We have chosen:
$$\zeta =10.$$

It is obvious that in this case $$M'_{SUSY}=100 \,\,\,{\mbox
{TeV}}.$$ $$M'_R=10^{12}\,\,\, {\mbox {GeV}},$$ which give:
$$M'_{GUT}= 2.46\cdot 10^{16}\,\,\, {\mbox {GeV}}. $$
Here $$M'_{SGUT}= 10^{17}\,\, {\mbox {GeV}}$$ is arbitrarily
chosen and gives: \be {\alpha'}_{SGUT'}^{-1}= 28.88. \lb{30x} \ee
Finally, we obtain the $E_6$ unification at the scale
$$ M'_{E_6} = M_{E_6}=5\cdot 10^{18} \,\,\, {\mbox {GeV}}, $$
where the inverse coupling constant attains the value
 \be
{\alpha'}_{E_6}^{-1} = \alpha_{E_6}^{-1}= 40.82\,.\lb{31x} \ee

Fig. 1 visually demonstrates the possibility of the $E_6$
unification in our model, with the breaking scheme $E_6\to SU(3)^3$
in both O- and M- worlds.

\section{Cosmological constant in the model of
shadow  theta-particles}

In the previous section we have presented an example of the gauge
coupling constant evolutions from the SM up to the $E_6$ unification
scale in the ordinary and mirror worlds with broken mirror parity.
We have assumed that the $E_6$ group of symmetry (inspired by
superstring theory) undergoes the breaking:
         $ E_6 \to SU(3)_C\times SU(3)_L\times SU(3)_R $
in both worlds (O and M) and gives the SM group of symmetry at lower
energies. Of course, such a Universe could exist, but it is
difficult to find a simple explanation why the observable $CC$ has
such a tiny value (\ref{2}), since, as we have discussed in
Subsection \ref{SM}, the considered model with mirror world does not
have an extremely large radius of confinement of any gauge
interaction. Thus, it is impossible to conceive a vacuum with
extremely small vacuum energy density.

\subsection{Theta-particles}

In Refs. \cite{33,34}, Okun developed a theory of $\theta$-particles
assuming that in Nature there exists the symmetry group (\ref{6}):
$$ SU(3)_C\times SU(2)_L\times U(1)_Y\times SU(2)_{\theta},$$
which contains a non-Abelian $SU(2)_{\theta}$ group with massless
gauge particles, ``thetons'', having a macroscopic confinement
radius $1/\Lambda_{\theta}$. Later, in Ref. \cite{45}, it was
assumed that if any $SU(2)$ group with the scale $\Lambda_2 \sim
10^{-3}$ eV exists, then it is possible to explain the small value
(\ref{2}) of the observable $CC$. The latter idea was taken up in
Refs. \cite{15,16,17,18,19,20}.

In the present context we can obtain the group of symmetry (\ref{6})
in the shadow world, but not in the ordinary world, as a natural
consequence of different schemes of the $E_6$-breaking in the O- and
Sh-worlds. $\theta-$particles are absent in the ordinary world,
because their existence is in disagreement with all experiments.
However, they can exist in the shadow world. By analogy with the
theory developed in \ct{33,34}, we consider shadow thetons
${\Theta'}^i_{\mu\nu}$, $i=1,2,3$, which belong to the adjoint
representation of the group $SU(2)'_{\theta}$, three generations of
shadow theta-quarks $q'_{\theta}$ and shadow leptons $l'_{\theta}$,
and the necessary theta-scalars $\phi'_{\theta}$ for the
corresponding breakings. Shadow thetons have macroscopic confinement
radius
 $1/\Lambda'_{\theta}$, and we assume that
\be \Lambda'_{\theta}\sim 10^{-3}\,\, {\mbox{eV}}. \lb{32x}   \ee

\subsection{Shadow World}

Superstring theory has led to the speculation that there may exist
another form of matter -- ``shadow matter'' -- in the Universe (see
\ct{29}), which only interacts with ordinary matter via gravity or
gravitational-strength interactions. The concept of Shadow Universe
was first introduced by K. Nishijima \cite{nishijima}. Further
development of this idea was given in Ref. \cite{okun-pomer} in
connection with neutrino experiments. The shadow world, in contrast
to the mirror world, can be described by another group of symmetry
(or by a chain of groups of symmetry), which is different from the
ordinary world symmetry group.

In our model, we shall adopt for the O-world the breaking
         $$E_6\to SO(10)\times U(1),$$
while for the Sh-world, given the fact that at low energies we wish
to have the extra $SU(2)_\theta'$ group, we shall consider the
breaking
$$E'_6\to SU(6)'\times SU(2)'. $$

\subsection{The breaking ${E_6 \to SO(10)\times U(1)_Z}$ in the
ordinary world}

Let us consider now the evolutions of the inverse coupling constants
in the O-world and in the Sh-world, with the values of parameters
$\zeta$ and $\xi$ fixed to
\be \zeta= 30 \quad \rm{and} \quad
                         \xi= 1.5. \lb{33} \ee
As in the first part of our work, we again consider the running of
all gauge coupling constants in the SM and its extensions which are
well described by the one-loop approximation of RGEs. We assume that
in the ordinary world, from the SM up to the $E_6$ unification,
there exists the following chain of symmetry groups:

$$SU(3)_C\times
SU(2)_L\times U(1)_Y \to  [SU(3)_C\times SU(2)_L\times
U(1)_Y]_{{SUSY}}
$$ $$\to
 SU(3)_C\times SU(2)_L \times SU(2)_R\times U(1)_X\times
 U(1)_Z$$
 $$ \to SU(4)_C\times SU(2)_L \times SU(2)_R\times U(1)_Z $$ \be \to
 SO(10)\times U(1)_Z \to E_6. \lb{34} \ee

\subsubsection{Standard Model and MSSM}

Starting with the SM and MSSM in the ordinary world we repeat the
results of Subsection 4.2. The running of the inverse coupling
constants as functions of $x=\log_{10}\mu$ is presented in Fig.~2
(a,b), using the scales $ M_{{SUSY}}= 10\,\,\rm{TeV}$ and $M_R
=2.5\cdot 10^{14}$ GeV.  In these figures, solid lines correspond to
the ordinary world. Fig.~2(b) shows the running of the gauge
coupling constants near the scale of the $E_6$ unification (for
$x\ge 15$).

\subsubsection{ Left-right symmetry, $SO(10)$ and $E_6$ unification}

We assume that the following supersymmetric left-right symmetry
originates at the seesaw scale $M_R$ \ct{42,43,44} :
$$
 SU(3)_C\times SU(2)_L\times
SU(2)_R\times U(1)_X\times U(1)_Z. $$
At the next step, we assume that the group $$ SU(4)_C\times
SU(2)_L\times SU(2)_R$$ of the Pati-Salam model \ct{42} originates
at the scale $M_4$, giving the following extension of the symmetry
group :
\be
 SU(3)_C\times SU(2)_L\times
SU(2)_R\times U(1)_X\times U(1)_Z$$ $$ \to SU(4)_C\times
SU(2)_L\times SU(2)_R\times U(1)_Z.  \lb{11z}\ee
At the scale $M_{{GUT}}$, the $ SO(10)$-unification occurs:
\be
        SU(4)_C\times SU(2)_L \times SU(2)_R \to SO(10).  \lb{12z} \ee
The evolution of $\alpha_{10}^{-1}(\mu)$, corresponding to $
SO(10)$, occurs from the scale $ M_{{GUT}}$ up to the super-GUT
scale $ M_{{SGUT}}$ of the $E_6$ unification:
$$ SO(10)\times U(1)_Z \to E_6. $$
The super-GUT scale is \footnote{A comment on terminology: the
scales $M_{GUT}$ and $M_{SGUT}$ in the model with Sh-world have
nothing to do with the scales considered in the first part of the
work, i.e. in the model with M-world (Section 4). }
$$M_{{SGUT}}= M_{{E_6}}\sim 10^{18}
{\mbox{GeV}}.$$
The coefficients (slopes) $b_i$, describing the running of the
coupling constants with our choice of gauge groups and particle
content in the O-world, are given in Table 2 (in accord with Refs.
\cite{19,7f,38,39}).

\begin{table}
\begin{center}
{\scriptsize
\begin{tabular}{|r|c|c|c|c|}\hline
NonSUSY groups:&$SU(3)_C$&$SU(2)_L$   &$U(1)_Y$               &           \\
         $b_i$:&$7$      &$19/6$      &$-41/10$               &           \\\hline

   SUSY groups:&$SU(3)_C$&$SU(2)_{L,R}$&$SU(2)_L\times SU(2)_R$&$U(1)_Y$    \\
   $b_i^{SUSY}$:&$3$      &$-1$        &$b_{22}= -2$           &$-33/5$     \\

               &         &            &                       &           \\
               &$SU(4)$  &$U(1)_X$    &$U(1)_Z$               &$SO(10)$   \\
               &$b_4=5$  &$-33/5 $    &$-9$                   &$b_{10}=1$  \\\hline
\end{tabular}\caption{The coefficients $b_i$ in the O-world with the breaking $E_6\to SO(10)\times U(1)_Z$.}
}
\end{center}
\end{table}

\subsection{Gauge coupling constant evolutions in the shadow world}

\subsubsection{Gauge coupling constant evolutions in the shadow
SM$'$ and  MSSM$'$}

Let us consider now the shadow world and the extension of the SM$'$,
for the moment ignoring the extra $SU(2)'$ group which survives at
low energies in this scheme of breaking. The first steps of such an
extension are:
$$SU(3)'_C\times
SU(2)'_L\times U(1)'_Y$$
$$\to [SU(3)'_C\times
SU(2)'_L\times U(1)'_Y ]_{SUSY}$$ $$ \to [SU(3)'_C\times
SU(2)'_L\times U(1)'_X\times U(1)'_Z]_{SUSY},
$$
and then
$$[SU(3)'_C\times
SU(2)'_L\times U(1)'_X]_{SUSY}$$ $$\to  SU(4)'_C\times SU(2)'_L.$$
In the SM$'$-sector of the shadow world we have the following
evolutions: \be
 {\alpha'}_i^{-1}(\mu) = {\alpha'}_i^{-1}(M'_t) + \frac{b_i}{2\pi}t'
 = \frac{b_i}{2\pi}\ln \frac{\mu}{\Lambda'_i}   \lb{21z} \ee
for $i=1,2,3$, where $M'_t =\zeta M_t.$ We should point out that the
scales $\Lambda'_i$ and $\Lambda_i$ are different, though the slopes
are the same:
$$b'_i = b_i.$$

As in the mirror world of the first part of our paper, the
supersymmetry breaking scale in the Sh-world is larger: $
M'_{{SUSY}} = \zeta M_{{SUSY}}$. The shadow MSSM$'$ leads to the
evolutions $ {\alpha'}_i^{-1}(\mu)$ (where $i=1,2,3,$), which run
from the scale $M'_{{SUSY}}$ up to the scale $M'_R$ in the Sh-world.

At the scale $M'_R = M_R$ the shadow right-handed neutrinos appear
and the chain of possible symmetries leading to the $E'_6$
unification is (see Ref. \cite{19}):
\be [SU(3)'_C\times SU(2)'_L\times U(1)'_Y]_{{SUSY}}$$
$$\to [SU(3)'_C\times SU(2)'_L\times U(1)'_X\times
U(1)'_Z]_{{SUSY}}, \lb{26} \ee \be [SU(3)'_C\times SU(2)'_L\times
U(1)'_X]_{{SUSY}}\to SU(4)'_C\times SU(2)'_L. \lb{27z} \ee The
coefficients $b_i$ (slopes), describing the evolutions in the shadow
world, are given in Table 3 (see Refs. \cite{19,7f,38,39}).

\begin{table}
\begin{center}
{\scriptsize
\begin{tabular}{|r|c|c|c|c|}\hline
Nonsupersymmetric groups: & $\large  SU(3)'_C$ & $\Large  SU(2)'_L$ & $\Large  SU(2)'_{\theta }$ & $\Large  U(1)'_Y$ \\
                     $\Large  b_i$: &     $7$      &    $19/6$    &    $3$    &  $-41/10$   \\\hline

  Supersymmetric groups: & $\Large  SU(3)'_C$ & $\Large  SU(2)'_L$ & $\Large  SU(2)'_{\theta}$ & $\Large  U(1)'_Y$ \\
                     $\Large  b_i$: &     $3$      &    $-1$      &     $-2$      &  $-33/5$    \\
                            &            &            &            &           \\
                            & $\Large  SU(4)'_C$  & $\Large  U(1)'_X$  & $\Large  U(1)'_Z$  & $\Large  SU'(6)$  \\
                            &     $5$      &    $-33/5$   &    $-9$      &   $11$      \\\hline
\end{tabular}
} \caption{ The coefficients $b_i$ in the Shadow World.}
\end{center}
\end{table}

\subsubsection{Shadow gauge coupling constant evolutions from
$SU(6)'$}

In the shadow world the evolutions are quite different from the
O-world. As a result, at the GUT scale $M'_{{GUT}}$ we reach
$SU(6)'$-unification, and not $SO(10)'$-unification:
\be SU(4)'_C\times SU(2)'_L\times U(1)'_Z  \to SU(6)'. \lb{28z} \ee
Then the $SU(6)'$ evolution occurs in the Sh-world up to the
super-GUT-scale
$$M'_{{SGUT}} = M'_{{E_6}}.$$
In the Sh-world the final chain is: \be SU(6)'\times SU(2)'\to E'_6,
\lb{29z} \ee
where the $SU(2)'$ survives unbroken up to the low energies of
SM$'$.

Now we are confronted with the question: what group of symmetry
$SU(2)'$, unknown in the O-world, exists in the Sh-world, ensuring
the $E'_6$ unification at the super-GUT-scale
$M'_{SGUT}=M'_{E_6}=M_{E_6}$? In this work we assume that this new
$SU(2)'$ group is precisely the $SU(2)'_{\theta}$ gauge group of
symmetry suggested in Refs. \cite{33,34}.

The unification $E_6=E'_6$ occurs  at the scale :
\be M'_{{SGUT}} = M'_{{E_6}} = M_{{SGUT}} = M_{{E_6}}\simeq
10^{18}\,\,{\mbox{GeV}} \lb{30z} \ee
and restores the mirror parity MP.

Finally, we obtain the following chain of symmetry breakings in the
shadow world:
$$
E'_6 \to SU(6)'\times SU(2)'_{\theta} \to SU(4)'_C\times
SU(2)'_L\times SU(2)'_{\theta}\times U(1)'_Z $$
$$ \to SU(3)'_C\times
SU(2)'_L\times SU(2)'_{\theta}\times U(1)'_X \times U(1)'_Z
$$ \be \to SU(3)'_C\times SU(2)'_L\times SU(2)'_{\theta}\times
U(1)'_Y.
                                          \lb{31z} \ee

\subsection{New shadow gauge group $SU(2)'_{\theta}$}

The reason for our choice of the $SU(2)'_{\theta}$ group was to
obtain the evolution ${\alpha'}_{2\theta}^{-1}(\mu)$, which leads to
the new scale (\ref{32x}) in the shadow world at extremely low
energies, according to the ideas considered in
Refs.\ct{15,16,17,18,19,20,33,34}.

By comparison with the content of the 27-plet of $E_6$ having 16
fermions (see Eqs.~(9-11)), we should consider theta-quarks as
$\theta-$doublets and shadow leptons as $\theta-$singlets. Then we
have 12+4 fermions, with 12 quarks having $3\times 2\times 2$
degrees of freedom, corresponding to $SU(3)_C\times SU(2)_L\times
SU(2)_{\theta}$. The scalars $\phi'_{\theta}$ also can be considered
as doublets of $SU(2)'_{\theta}$. Theta-quarks can be heavier than
ordinary quarks, having additional interactions with thetons.

We start at high energies $\mu > M'_t$ with three generations of
theta-quarks and assume the existence of two doublets of scalar
fields $\phi'_{\theta}$ with $\langle\phi'_{\theta}\rangle\sim
10^{-3}$ eV. Then we have the following slopes given by Refs.
\cite{19},\ct{7f} (see also \ct{38} and \ct{39}):
\be
    b_{2\theta} = 3 \quad \rm{ and}\quad  b_{2\theta}^{SUSY}= -2.  \lb{32z} \ee
Of course, near the scale  $\Lambda'_{\theta}$ only theta-quarks of
the first generation contribute, and it is easy to obtain the value
$\Lambda'_{\theta}\approx 3\cdot 10^{-3}$ eV. Theta-quarks of the
first generation are stable, due to the conservation of theta-charge
\ct{33,34}.

We also consider a complex scalar field
 $$ \varphi_{\theta} = (1,1,0,1), $$
which is a singlet under the symmetry group $$G' = SU(3)'_C\times
SU(2)'_L\times U(1)'_Y\times SU(2)'_{\theta}.$$ This comes from
27-plet of the $E'_6$ unification (see Eq.~(14)).

In Figs.~3 (a,b) we have shown the evolutions of all
${\alpha'}^{-1}(\mu)$ in the Sh-world, given by dashed lines, together
with ${\alpha'}_{2\theta}^{-1}(\mu)$.

The comparison of the evolutions in the O- and Sh-worlds is
presented in Figs.~4 (a,b).

The parameters of our model are as follows:
$$M_{SUSY}=10\,\, {\mbox{TeV}},
$$
 $$\zeta =30.$$

In this case we have: $$M'_{SUSY}=300 \,\,\,{\mbox {TeV}},$$ and
$$M_R=M'_R=2.5\cdot 10^{14}\,\,\,{\mbox {GeV}}.$$
Here

$$M_{4}=9.40\cdot 10^{15}\,\,\, {\mbox {GeV}},$$
$$M'_{4}=3.01\cdot 10^{17}\,\,\, {\mbox {GeV}},$$

$$M_{GUT}=1.10\cdot 10^{16}\,\,\, {\mbox {GeV}},$$
$$M'_{GUT}=6.37\cdot 10^{17}\,\,\, {\mbox {GeV}},$$

$$M_{SGUT}=M'_{SGUT}=M_{E_6}=6.98\cdot 10^{17}\,\,\, {\mbox {GeV}},$$
and
\be \alpha_{E_6}^{-1}=27.64.\lb{A} \ee

\section{Cosmological Constant, Dark Energy and Dark Matter}

From the point of view of particle physics the cosmological
constant naturally arises as energy density of the vacuum.

For the present epoch, the Hubble parameter $H$ is given by the
following value:
\be H = 1.5 \times 10^{-42}\,\,{\rm{GeV}}, \lb{60} \ee
and the critical density of the Universe is
\be \rho_{c} = 3H^2/8\pi G = {(2.5\times 10^{-12}\,\,
{\rm{GeV}})}^4. \lb{61} \ee
According to the Particle Data Group \ct{41}, the fraction of the
dark energy corresponds to
\be \rho_{DE} \approx 0.75\ \rho_c \approx (2.3 \times 10^{-3}
\,\, {\rm eV})^4. \lb{62} \ee
The $\Lambda CDM$ cosmological model predicts that the
cosmological constant $\Lambda$ is equal to
$$ CC = \Lambda = \rho_{vac}= \rho_{DE}.$$ Given by Eq.(\ref{62}),
$CC$ is extremely small. This is a result of recent cosmological
observations (see, for example, Refs.~\ct{45a,45b,45c}).

Modern Quantum Field Theory (QFT) gives an energy scale of
$\Lambda$ much larger than the present cosmological value. This is
the cosmological constant problem \ct{45d} and was well known to
exist long before the discovery of the accelerated expansion of
the Universe in 1998.

There have been a number of attempts to solve this problem.

Previously in Ref.~\cite{46a} and also in Ref. \cite{46b,46c} it
was shown that SUGRA models which ensure the vanishing of the
vacuum energy density near the physical vacuum lead to a natural
realization of the Multiple Point Model (MPP) \ct{46d, 46e, 46f}
(see also the reviews \ct{46g,46k}) describing the degenerate
vacua with naturally tiny $CC$.

In the present paper it is assumed that all contributions to $CC$
 are canceled, except the condensates (zero mode
contributions) of shadow $\theta$ particles, especially thetons.
Their contributions provide the minimum of the overall effective
potential:
\be min\,\,V_{eff} = \rho_{DE} = \rho_{vac}\simeq
\Lambda'^4_\theta \simeq (2.3 \times 10^{-3} \,\, {\rm eV})^4.
\lb{63} \ee
It is essential that in the low energy region we have the $G_{SM}$
symmetry group in the O-world, but the $G'_{SM'}\times
SU(2)'_{\theta}$ group of symmetry in the Sh-world. If we assume
that superstring theory, or supergravity provides the cancellation
of the SM and SM' contributions to $CC$, then we can relate the
value (\ref{63}) with the result of confinement given by
$SU(2)'_{\theta}$. This phenomenon is not obvious, but it is
possible to fix the existence of the group of symmetry
$SU(2)'_{\theta}$ as a consequence of the $E_6$-breakdown in the
Sh-world, using (modern and future) astrophysical measurements. We
have obtained $E_6$ unification in the 4-dimensional space
considering the spontaneous compactification of the
ten-dimensional $E_8\times E'_8$ superstring theory. The breakdown
of compactification could be important in solving the cosmological
problem (see for example \ct{47b,47c,47d}). We hope to investigate
this problem in forthcoming communications.

There exists an axial $U(1)_A$ global symmetry in our theory with
a current having $SU(2)'_\theta$ anomaly, which is spontaneously
broken at the scale $f_{\theta}$ by a singlet complex scalar field
$\varphi_{\theta}$, with a VEV $\langle \varphi \rangle =
f_\theta$,\, i.e.
\be \varphi = (f_\theta + \sigma) \exp(i a_\theta/f_\theta).
\lb{52} \ee
The boson $a_{\theta}$ (imaginary part of the singlet scalar field
$\varphi_{\theta}$) is an axion and could be identified with a
massless Nambu-Goldstone (NG) boson if the $U(1)_A$ symmetry is
not spontaneously broken. However, the spontaneous breaking of the
global $U(1)_A$ by $SU(2)'_{\theta}$ instantons inverts
$a_{\theta}$ into a pseudo Nambu-Goldstone (PNG) boson.

The singlet complex scalar field $\varphi_{\theta}$ reproduces a
Peccei-Quinn (PQ) model \ct{47} (well known in QCD, but having a
different meaning in our model). In the shadow world with shadow
$\theta$-particles the vacuum energy density is given by
Eq.~(\ref{63}), which means that
\be \Lambda'_\theta \approx 2.3 \times 10^{-3} \,\,{\rm{eV}}.
\lb{67} \ee
Near the vacuum, a PNG mode $a_\theta$ emerges the following PQ
axion potential:
\be V_{PQ}(a_\theta) \approx {(\Lambda'_\theta)}^4
         \left(1 - \cos(a_\theta/f_\theta)\right).  \lb{53} \ee
This axion potential exhibits minima at
\be \cos(a_{\theta}/f_\theta) = 1, \lb{54} \ee
i.e.
\be {(a_{\theta})}_{min}= a_n = 2\pi n f_\theta, \quad n = 0,1,...
\lb{55} \ee
For small fields $a_\theta$ we expand the effective potential near
the minimum:
\be    V_{eff} \approx (\Lambda'_\theta)^4 (1 +
            \frac 12 (a_\theta/f_\theta)^2 + ...) = {(\Lambda'_\theta)}^4
            + \frac 12 m^2 a_\theta^2 + ...,  \lb{56} \ee
 and hence the PNG axion mass squared is given by:
\be m^2\sim {\Lambda'_{\theta}}^4/f^2_{\theta}.  \lb{57} \ee
Let us assume that at the cosmological epoch when $U(1)_A$ was
spontaneously broken, the value of the axion field $a_\theta$ was
deviated from zero, and it was $a_{\theta,in}\sim f_\theta$. The
value of the scale $f_\theta \sim 10^{18}$ GeV (near the $E_6$
unification breaking scale) makes it natural that the $U(1)_A$
symmetry was broken before inflation, and the initial value
$a_{\theta,in}$ was inflated above the present horizon. So after
the inflation breaking scale, and in particular in the present
Universe, the field $a_\theta$ is spatially homogeneous
(constant), and the initial energy density corresponding to
$a_{\theta,in}$ is also spatially homogeneous:
 \be \rho_{in} = V(a_{\theta,in}) \backsimeq
\Lambda'^4_\theta \left(1 - \cos(a_{\theta,in}/f_\theta)\right),
\lb{58} \ee
and its value changes only with time.

For the expanding Universe the equation of motion (EOM) of the
classical field $a_\theta$ is:
\be \frac{d^2 a_\theta}{dt^2} + 3 H \frac{d a_\theta}{dt} +
V'(a_\theta) = 0, \lb{59} \ee
where $H$ is the Hubble parameter (\ref{60}).

For small $a_\theta$  we have:
\be V'(a_\theta)= m^2 a_\theta. \lb{64} \ee
If $\Lambda'_\theta \sim 10^{-3}$ eV and $f_\theta \sim 10^{18}$
GeV, then from Eq.~(\ref{57}) we obtain the value of the axion
mass:
\be m\sim {\Lambda'_{\theta}}^2/f_{\theta}\sim 10^{-42}
\,\,{\rm{GeV}}. \lb{65} \ee
Now, it is natural to assume that the initial velocity
$\dot{a}_{\theta,in}$ was small:
\be \dot{a}_{\theta,in}\sim H f_\theta. \lb{66} \ee
Then, for $3H^2 \gg m^2$  the potential curvature $V'(a_\theta)$
in the above EOM can be neglected, and we have a solution with
$a_\theta$ remaining the constant in time.

Now, having $m^2 < 3H^2$, we see that the classical PNG field
$a_\theta$ does not start the oscillation and in the present epoch
its energy density remains constant (does not scale with time). In
this case, for the present epoch, the energy of the PNG field
$a_\theta$ can imitate dark energy, providing the equation of
state $\rho = w p$ with $w \approx -1$, but not exactly equal to
$-1$ (quintessence model). Of course, to claim that this can
explain the present amount of dark energy, one again must assume
that the major constant contributions to the cosmological term are
canceled by some means, i.e. the true cosmological constant is
almost zero due to some (yet unknown) symmetry (see for example
\ct{46c}), or due to dynamical reasons. Also the gravity itself
can be modified so that it does not feel the truly constant terms
in the energy (see for example Ref.~\ct{47a}). In this case one
can ascribe the present acceleration of the Universe to such a PNG
quintessence field, with the implication that the acceleration
will not be forever, but it will finish as soon as $m^2 \sim 3H^2$
will be achieved. After that the PQ classical energy will behave
as a dark matter component and not as dark energy.

In the present paper we have suggested a model in which our
Universe was trapped in the vacuum (\ref{63}) and exists there at
the present time with a tiny cosmological constant $CC$:
\be CC=(\Lambda'_\theta)^4\approx (2.3\times
10^{-3}\,\,\rm{eV})^4. \lb{67} \ee
Such properties of the present axion lead to the `$\Lambda CDM$'
model of our accelerating expanding Universe \ct{6,7,8,9,10}. By
this reason, the axion $a_{\theta}$ could be called an
`acceleron', and the field $\sigma$ given by Eq.~(\ref{52}) is an
inflaton.

\subsection{Dark matter}

The existence of dark matter in the Universe, which is
non-luminous and non-absorbing matter, is now well established in
astrophysics. Recently very interesting investigations of DM were
presented in Refs. \ct{48,49,50,51,52,53,54}.

For the ratios of densities,
\begin{equation}
\Omega_X = \rho_x/\rho_c,   \end{equation}
where $\rho_c$ is the critical energy density, cosmological
measurements give the following density ratios of the total
Universe \ct{41}:
\begin{equation}
\Omega_0 = \Omega_r + \Omega_M + \Omega_\Lambda = 1,
\end{equation}
where $ \Omega_r$ is a relativistic (radiation) density ratio and
$$
\Omega_{\Lambda} =\Omega_{DE}\sim 75\%\,,$$
for the mysterious Dark Energy, which is responsible for the
acceleration of the Universe, while
\begin{equation}
 \Omega_M \approx \Omega_B + \Omega_{DM} \sim 25\%,
 \end{equation}
with
$$\Omega_B \approx 4\%$$
for (visible) baryons and
$$\Omega_{DM} \approx 21\%$$
for the Dark Matter.

Here we propose that a plausible candidate for DM is a shadow
world with its shadow quarks, leptons, bosons and super-partners,
of which the shadow baryons are dominant:
 $$\Omega_{DM} \approx \Omega_{B'}.$$
We see that
$$\Omega_{B'} \approx 5\Omega_{B},$$
meaning that the shadow baryon density is larger than the ordinary
one.

The new gauge group $SU(2)'_{\theta}$ gives the running of
${(\alpha')}_{2\theta}^{-1}(\mu)$. Near the scale
$\Lambda'_{\theta}\sim 10^{-3}$ eV, the coupling constant
$g'_{2\theta}$ grows infinitely. But at higher energies (see
Figs.~3 and 4) this coupling constant is comparable with the
electromagnetic one. Here we would like to emphasize that the
shadow quarks $q'_{\theta}$ of the first generation are stable,
and can participate in the formation of shadow ``hadrons'',  which
can be considered as good candidates for the Cold Dark Matter
(CDM). So we have two types of shadow baryons: baryons $ b'$
constructed from shadow quarks $q'$ which are singlets of
$SU(2)'_{\theta}$, and baryons $b'_\theta$ constructed from the
quark $q'$ and two shadow $\theta$-quarks $q'_\theta$, in order to
preserve $\theta$-charge conservation. Then,
$$\Omega_{B'} = \Omega_{b'+b'_\theta}\approx 5\Omega_{B}.$$
We shall study in detail the DM in a forthcoming communication.

\section{Conclusions}

We have considered cosmological implications of the parallel
ordinary and mirror or shadow worlds, with broken mirror parity.
The parameter characterizing the breaking of MP is $\zeta = v'/v$,
where $v'$ and $v$ are the VEVs of the Higgs bosons in the M (or
Sh)- and O-worlds, respectively.

We have assumed that at very high energies there exists the $E_6$
unification predicted by superstring theory, which restores the
broken mirror parity MP at the scale $\sim 10^{18}$ GeV. We have
chosen a model which leads to asymptotically free $E_6$
unification, what is not always fulfilled.


In the first part of this paper, we have considered $E_6$
unification in the O- and M-worlds with the breaking $E_6 \to
SU(3)_C\times SU(3)_L\times SU(3)_R$. The model of unification is
very simple, but the unification is non-trivial. This breaking
scheme is the only one which enables us to obtain the $E_6$
unification of the O- and M- worlds below the Planck scale and
with plausible values for the SUSY and seesaw scales. The other
two possible breaking schemes of $E_6$ in the O- and M- world,
with broken MP, would not lead to unification.

Aiming at explaining the tiny value of the cosmological constant
$CC$, we have assumed the existence of a shadow (not mirror) world
parallel to our ordinary world. For numerical calculations, we
have used the value $\zeta= 30$. The comparison of the coupling
constants evolutions in the O- and Sh-worlds is given in Figs.~4
(a,b). The breaking of the unification $E'_6$ in the shadow world
is based on the group $E'_6\to SU(6)'\times SU(2)'_{\theta}$. The
$SU(2)'_{\theta}$ part of our model follows the theory of Okun
\ct{33,34} for theta-particles.

The existence of the new gauge group $ SU(2)'_{\theta}$ in the
Sh-world gives significant consequences for cosmology: it explains
the tiny value of $CC$ and the $\Lambda CDM$ model of our
accelerating Universe. It was shown in Subsection 6.1 that the
existence of the scale $\Lambda'_{\theta}\sim 10^{-3}$ eV explains
the value of cosmological constant, $CC\approx (2.3\times
10^{-3}\,\,\rm{ eV})^4,$ which is given by recent astrophysical
measurements. The bound states -- shadow ``hadrons'', the result
of confinement of shadow quarks $q'$ and $q'_{\theta}$ -- are
candidates for the Cold Dark Matter (CDM).

It should be emphasized that the present work opens the
possibility to study in detail the DM, and specify a grand
unification group, such as $E_6$, from Cosmology.

\section*{Acknowledgments}
Our deep thanks go to Zurab Berezhiani, Masud Chaichian, Maxim
Khlopov, Archil Kobakhidze, Holger Nielsen and Zhi-Zhong Xing for
useful discussions and advices. The support of the Academy of
Finland under the Projects Nos. 121720 and 127626 is greatly
acknowledged. L.V.L. thanks RFBR grant 09-02-08215-3. C.R.D.
gratefully acknowledges a scholarship from Funda\c{c}\~{a}o para a
Ci\^{e}ncia e Tecnologia ref. SFRH/BPD/41091/2007.

\clearpage\newpage \bfi \centering
\includegraphics[height=100mm,keepaspectratio=true,angle=0]{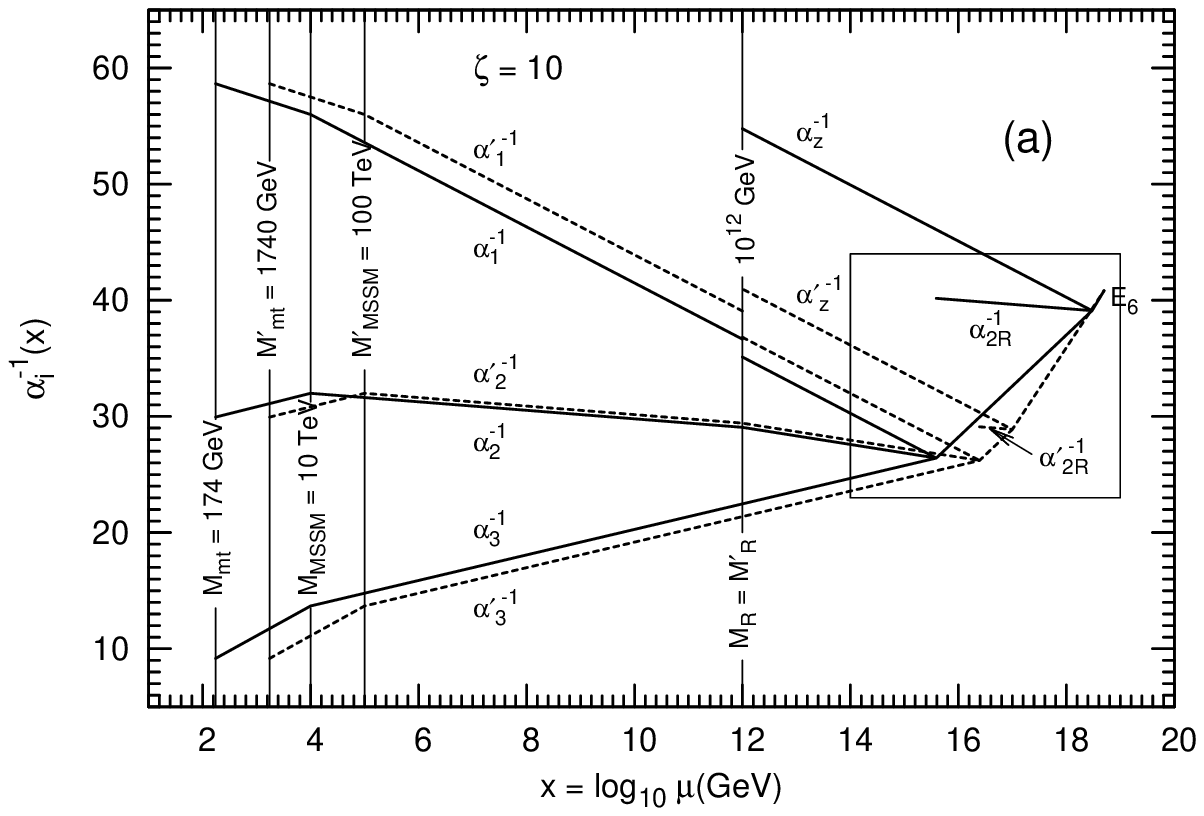}
\includegraphics[height=100mm,keepaspectratio=true,angle=0]{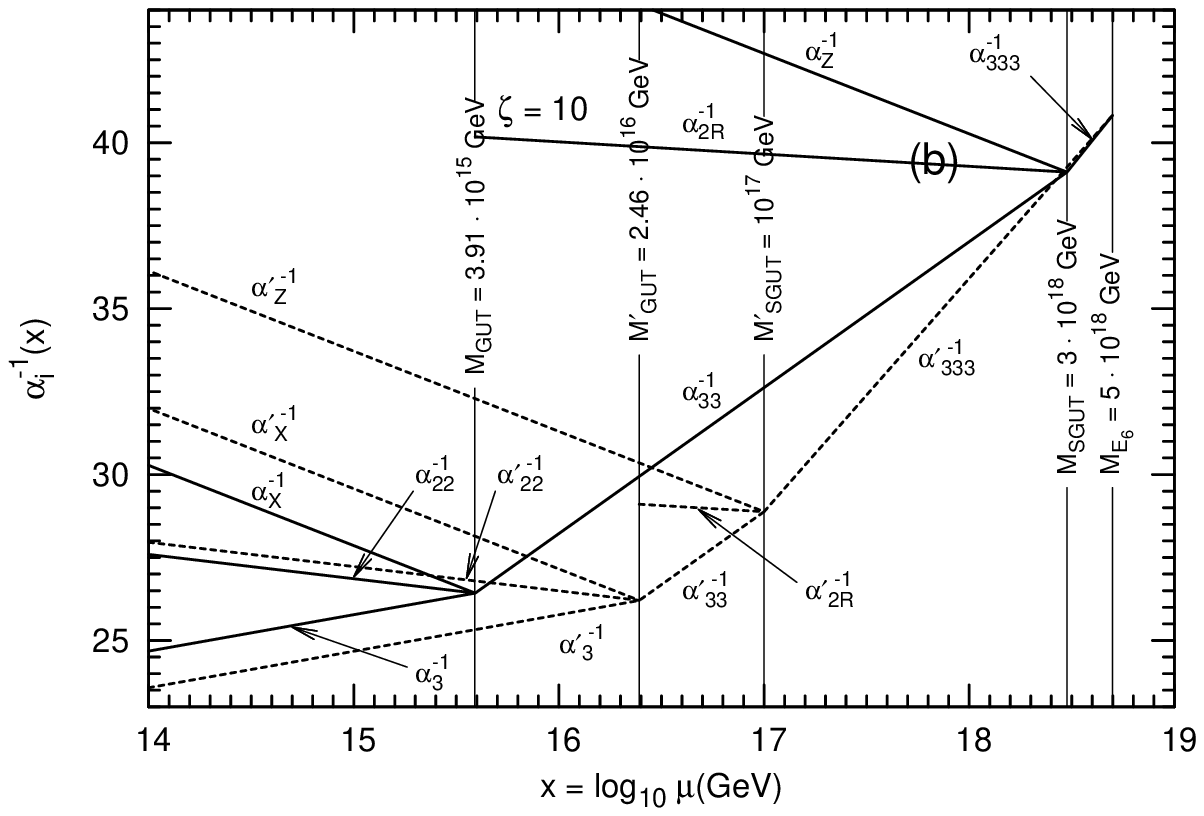}
\caption {The running of the inverse coupling constants
$\alpha_i^{-1}(x)$ in both ordinary and mirror worlds with broken
mirror parity, from the Standard Model up to the $E_6$ unification,
for SUSY breaking scales $M_{SUSY}= 10$ TeV, $M'_{SUSY}= 100$ TeV
and seesaw scales $M_R=M'_R=10^{12}$ GeV; $\zeta =
10$. This case gives: $M_{SGUT}=3\cdot 10^{18}$ GeV,
$M'_{SGUT}=10^{17}$ GeV, $M_{E_6}=5\cdot 10^{18}$ GeV and
$\alpha_{E_6}^{-1}=40.82$. Solid lines correspond to the ordinary
world, while dashed lines correspond to the mirror world.}\efi

\clearpage\newpage \bfi \centering
\includegraphics[height=100mm,keepaspectratio=true,angle=0]{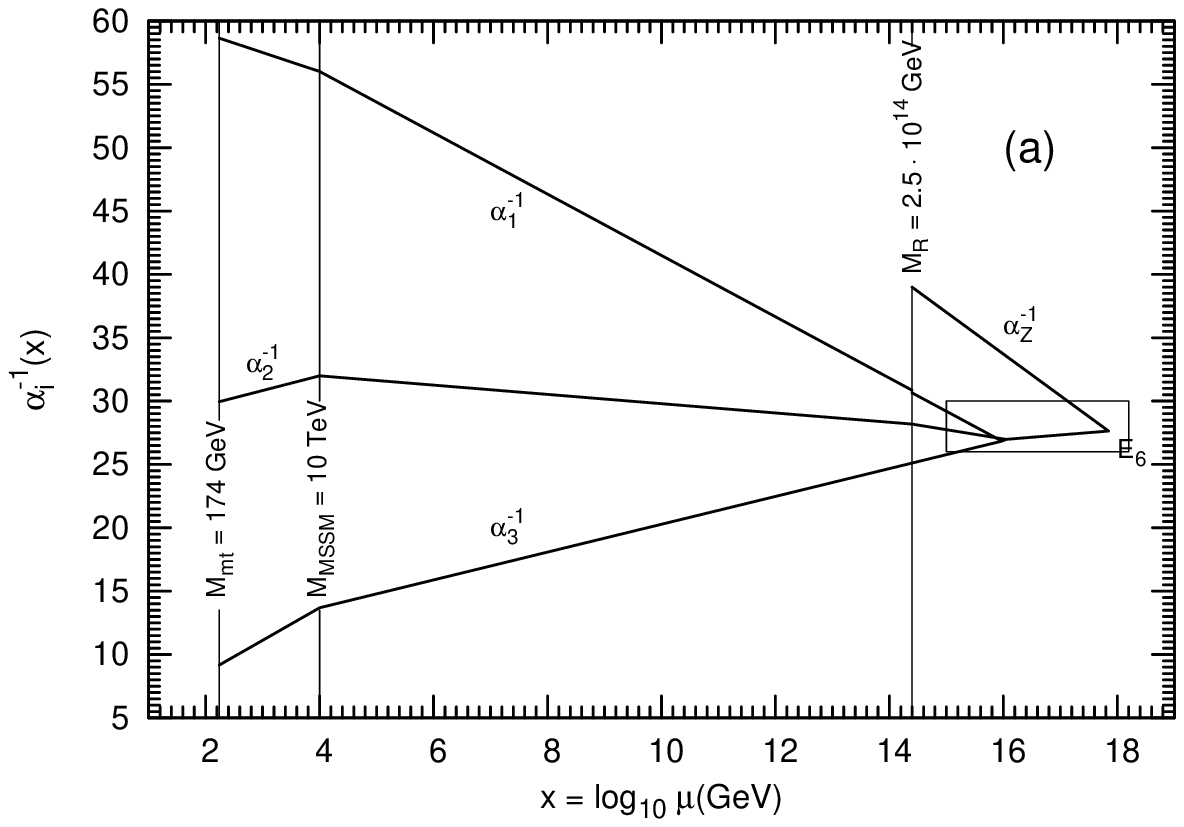}
\includegraphics[height=100mm,keepaspectratio=true,angle=0]{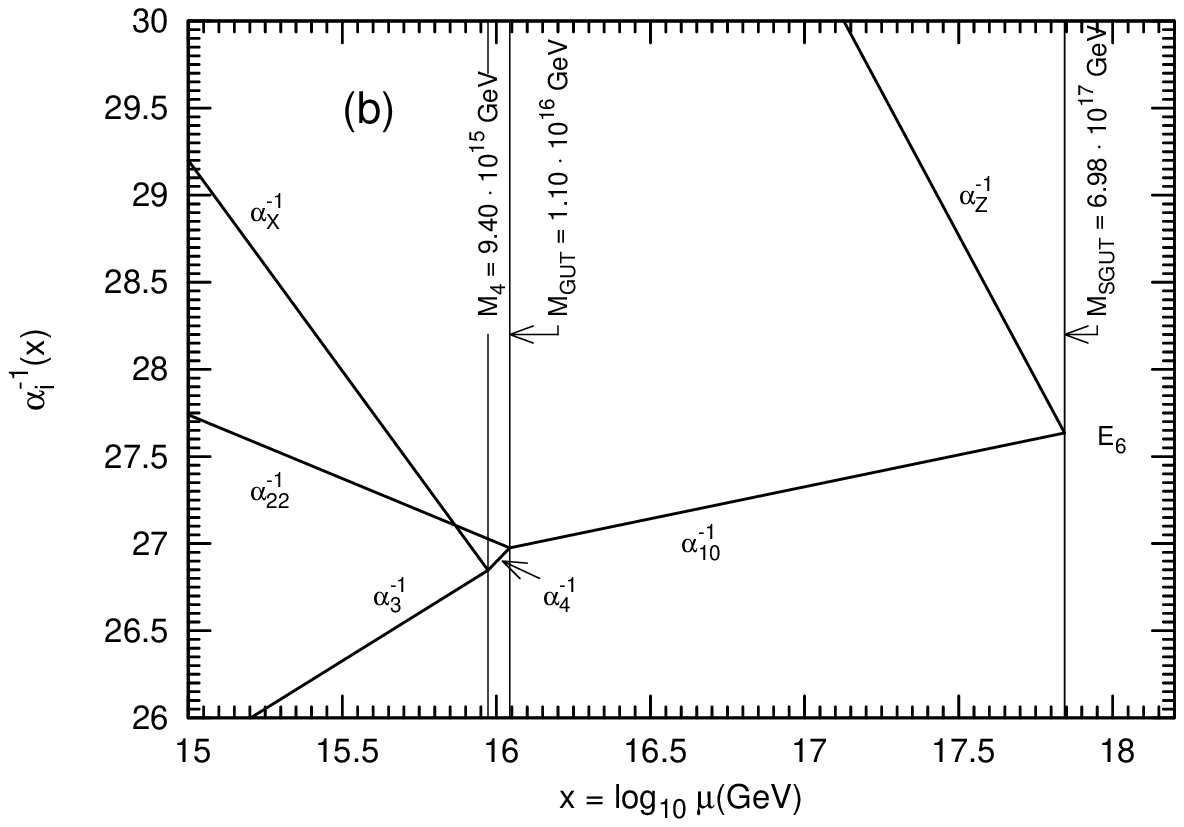}
\caption {Figure (a) presents the running of the inverse coupling
constants $\alpha_i^{-1}(x)$ in the ordinary world, from the
Standard Model up to the $E_6$ unification for SUSY breaking scale
$M_{SUSY}= 10$ TeV and seesaw scale $M_R=2.5\cdot 10^{14}$ GeV. This
case gives: $M_{SGUT}=M_{E_6}=6.98\cdot 10^{17}$ GeV and
$\alpha_{E_6}^{-1}=27.64$. Figure (b) is the same as (a), but zoomed
in the scale region from $10^{15}$ GeV up to the $E_6$ unification
to show the details.}\efi

\clearpage\newpage \bfi \centering
\includegraphics[height=100mm,keepaspectratio=true,angle=0]{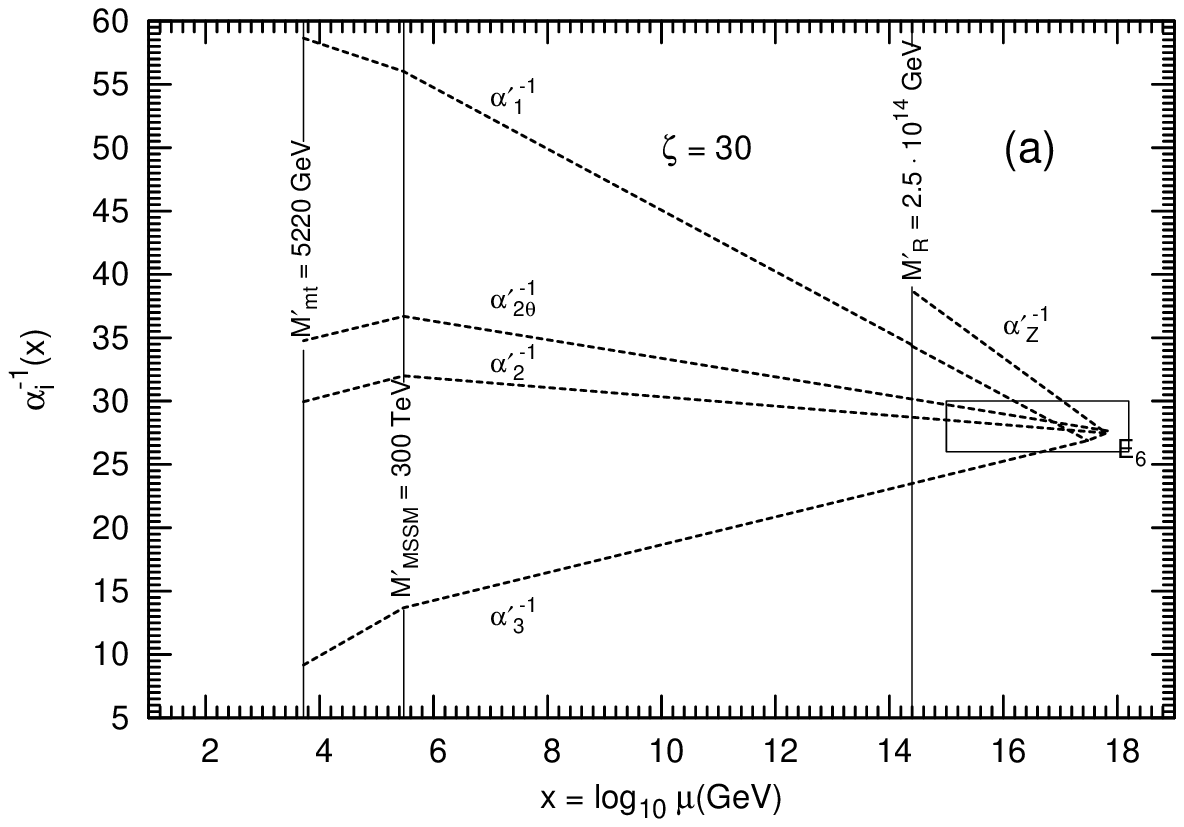}
\includegraphics[height=100mm,keepaspectratio=true,angle=0]{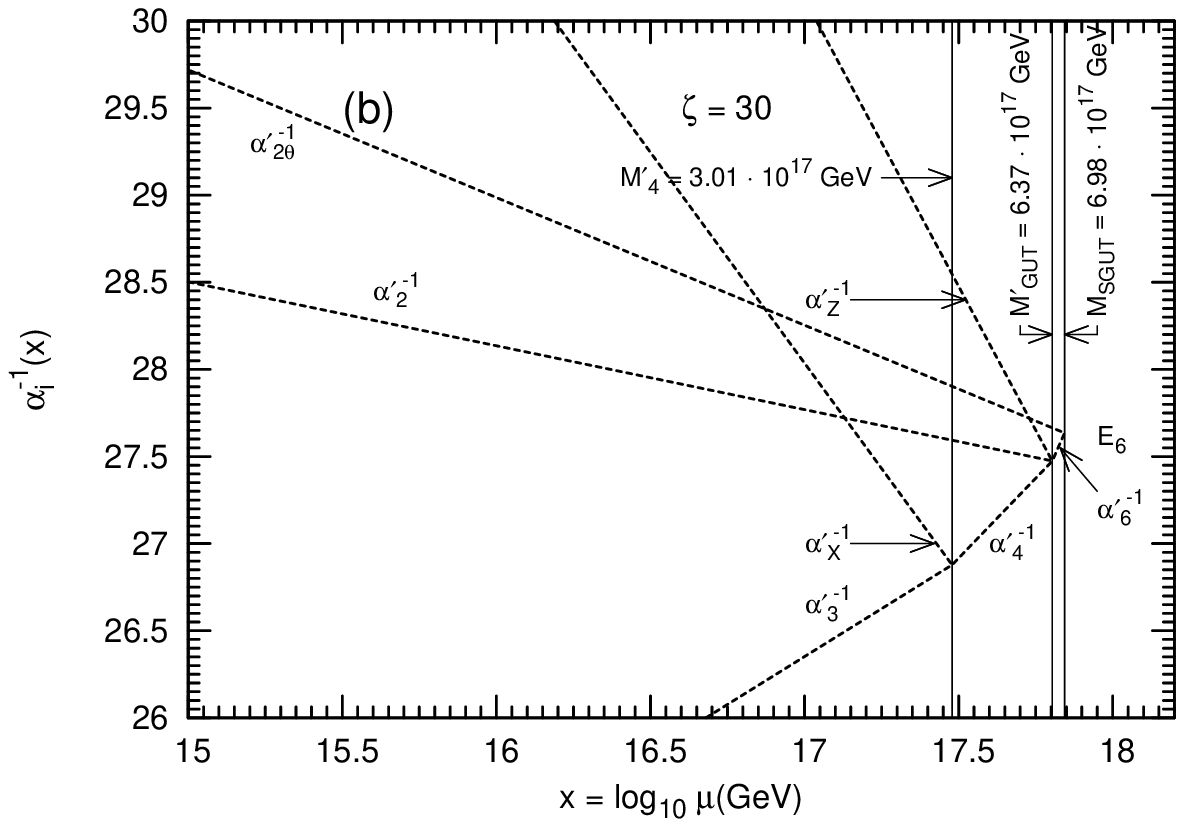}
\caption {Figure (a) presents the running of the inverse coupling
constants $\alpha_i^{-1}(x)$ in the mirror world from the Standard
Model up to the $E_6$ unification for mirror SUSY breaking scale
$M'_{SUSY}= 300$ TeV and mirror seesaw scale $M'_R=2.5\cdot
10^{14}$ GeV; $\zeta = 30$. This case gives: $M'_{SGUT}=M_{E_6}=6.98\cdot
10^{17}$ GeV and $\alpha_{E_6}^{-1}=27.64$. Figure (b) is the same
as (a), but zoomed in the scale region from $10^{15}$ GeV up to
the $E_6$ unification to show the details.} \efi

\clearpage\newpage \bfi \centering
\includegraphics[height=100mm,keepaspectratio=true,angle=0]{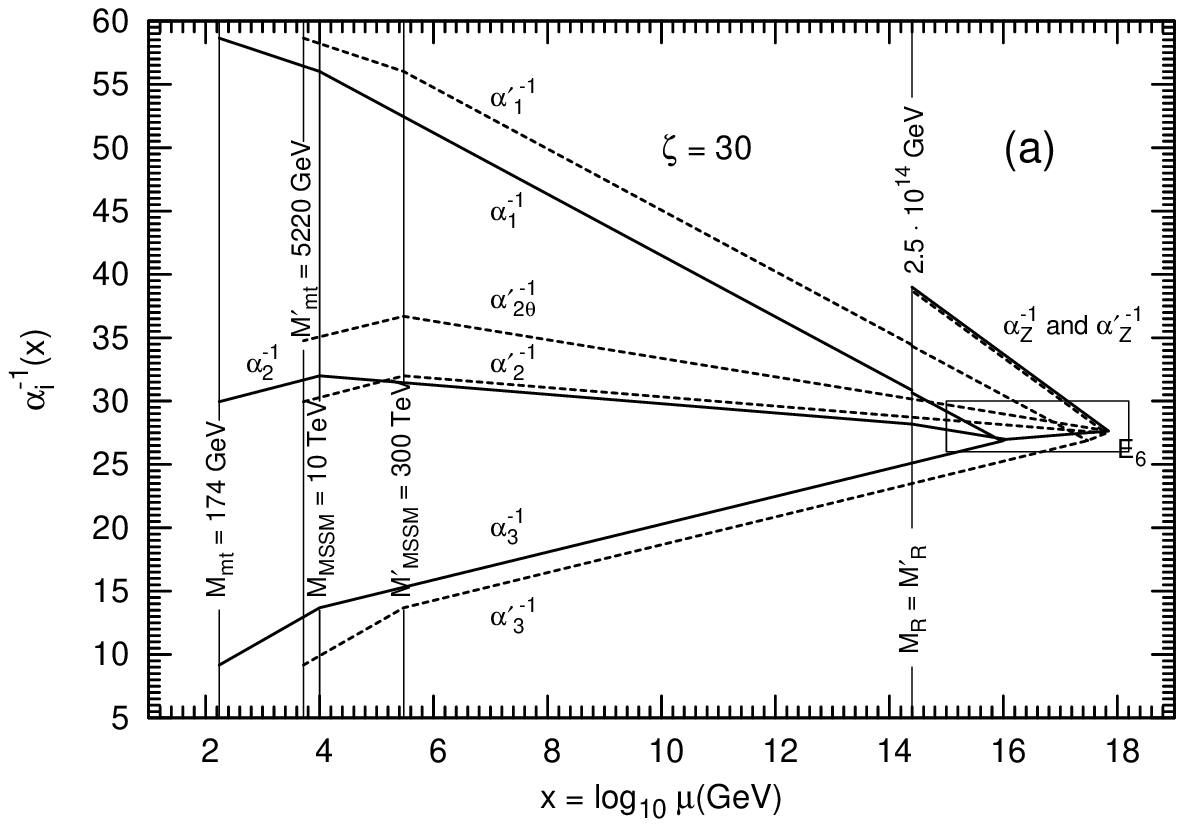}
\includegraphics[height=100mm,keepaspectratio=true,angle=0]{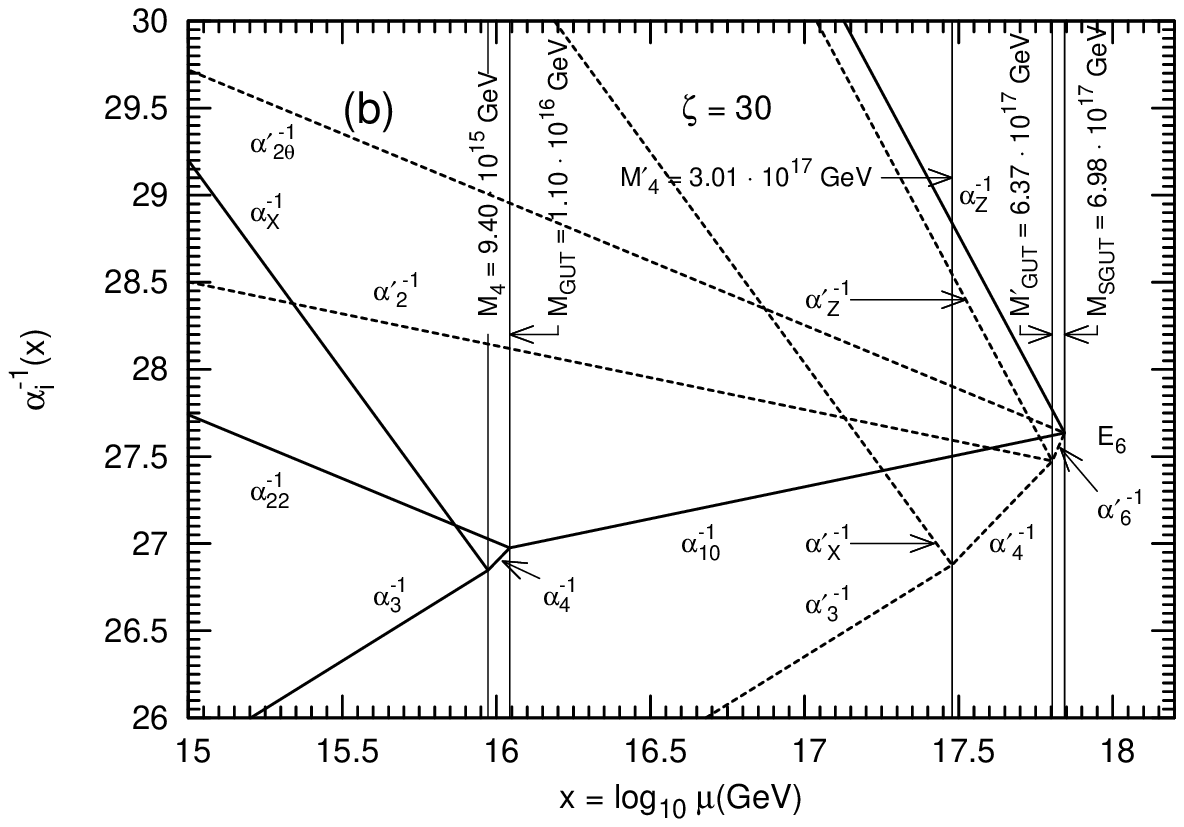}
\caption {Figure (a) shows the running of the inverse coupling
constants $\alpha_i^{-1}(x)$ in both ordinary and mirror worlds with
broken mirror parity, from the Standard Model up to the $E_6$
unification for SUSY breaking scales $M_{SUSY}= 10$ TeV, $M'_{SUSY}=
300$ TeV and seesaw scales $M_R=M'_R=2.5\cdot 10^{14}$ GeV; $\zeta = 30$. This case gives:
$M_{SGUT}=M'_{SGUT}=M_{E_6}=6.98\cdot 10^{17}$ GeV and $\alpha_{E_6}^{-1}=27.64$.
Figure (b) is the same as (a), but zoomed in the scale region from
$10^{15}$ GeV up to the $E_6$ unification to show the details.} \efi

\end{document}